\documentclass[12pt]{article}%
\usepackage{amsmath}
\usepackage{amsfonts}
\usepackage{amssymb}
\usepackage{graphicx}%
\providecommand{\U}[1]{\protect\rule{.1in}{.1in}}
\begin{document}

\title{Generalized Chern-Simons action principles for gravity}
\author{D C Robinson\\Mathematics Department\\King's College London\\Strand, London WC2R 2LS\\United Kingdom\\email: david.c.robinson@kcl.ac.uk}
\maketitle

\textbf{Abstract:} \ Generalized differential forms are employed to construct
generalized connections. \ Lorentzian four-metrics determined by certain of
these connections satisfy Einstein's vacuum field equations when the
connections are flat. \ Generalized Chern-Simons action principles with
Einstein's equations as Euler-Lagrange equations are constructed by using
these connections.

\section{Introduction}

Chern-Simons gravity and related topics have been the subject of extensive
investigation since the 1980's. \ Following pioneering papers such as
\cite{deser1}- \cite{witten1}, most of that research has dealt with gravity in
2+1 dimensions. \ In three dimensions source free general relativity, with or
without a cosmological constant, can be interpreted as Chern-Simons theories
of the relevant structure group and the field equations correspond to the
vanishing of the relevant curvature tensor. \ Reviews of that line of research
can be found in \cite{carlip} and a recent broad ranging discussion is given
in \cite{townsend}. \ Chern-Simons approaches to gravity in higher dimensions
have also been discussed, although to a lesser extent, as for example, in
\cite{jackiw}, and aspects of that work are reviewed in \cite{zanelli1} and
\cite{zanelli2}. \ In this paper a different approach is followed in that the
formalism of generalized forms is used to construct generalized Chern-Simons
actions for the four dimensional Einstein vacuum field equations with non-zero
cosmological constant. \ This type of approach, using generalized generalized
characteristic classes and generalized Chern-Simons forms, was initiated in
\cite{tung1} and \cite{tung2} and was subsequently developed in \cite{rob5}-
\cite{rob9}. \ The main new result in this paper is the construction of a
Palatini type Lagrangian for gravity from a generalized Chern-Simons integral
by using a generalized connection which is flat when the field equations are
satisfied. \ In order to do this results presented in \cite{rob9} are extended
from type $N=1$ to type $N=2$ generalized forms.

In sections\ two and three properties of type $N=2$ forms, generalized
connections and generalized Chern-Simons integrals are outlined. \ Much of the
material in these sections has been presented elswhere but it is included in
order to make this paper reasonably self-contained. \ In section four
attention is concentrated on type $N=2$ generalized connections with values in
the Lie algebras of $SO(p,q)$, where $p+q=5$. \ These connections are defined
on manifolds of dimension six or greater. \ When\ they are pulled back to
(boundary) four dimensional manifolds, and a regularity condition is
satisfied, they define Lorentian metrics there. \ In that case when the
generalized curvature of the $SO(p,q)$ connections vanishes these metrics
satisfy Einstein's vacuum field equations with non-zero cosmological constant.
\ The connections are also used to construct a parametrized family of
gravitational actions from generalized Chern-Simons integrals. \ These actions
have Einstein's equations as Euler-Lagrange equations.

It will be assumed that all geometrical objects are smooth and $M$ is an
$n-$dimensional real, smooth, orientable and oriented manifold. \ Bold-face
Roman letters are used to denote generalized forms and\ ordinary forms on $M$
are usually denoted by Greek letters. \ Sometimes the degree of a form is
indicated above it. \ The exterior product of any two forms, for example
$\alpha$ and $\beta,$ is written $\alpha\beta$, and as usual, any ordinary
$p-$form $\overset{p}{\alpha}$, with $p$ either negative or greater than $n$,
is zero. \ The Einstein summation convention is used.

\section{Type N=2 generalized differential forms}

In this section the properties of type $N=2$ differential forms on an $n$
dimensional manifold $M$ that are needed in this paper are reviewed. \ The
notation of \cite{rob5} and \cite{rob6} is again used. \ Further discussion of
type $N=2$ forms can be found in \cite{rob3} and \cite{rob4}.

Type $N=2$ generalized forms constitute a module $\Lambda_{(2)}^{\bullet
}=\Sigma_{p=-2}^{n}$ $\Lambda_{(2)}^{p}$ and obey the same algebraic and
differential relations as ordinary forms. \ In particular if
$\overset{p}{\mathbf{r}}$and $\overset{q}{\mathbf{s}}$ $\in\Lambda_{(2)}^{p}$
and $\Lambda_{(2)}^{q}$ are respectively a $p-$form and a $q-$form, then
$\overset{p}{\mathbf{r}}$ $\overset{q}{\mathbf{s}}=(-1)^{pq}%
\overset{q}{\mathbf{s}}\overset{p}{\mathbf{r}}$. \ A basis for type $N=2$
generalized forms consists of any basis for ordinary forms on $M$ augmented by
a pair of linearly independent minus one-forms $\{\mathbf{m}^{i}\}$
($i,j=1,2)$. \ Minus one-forms have the algebraic properties of ordinary
exterior forms but are assigned a degree of minus one. \ They satisfy the
ordinary distributive and associative laws of exterior algebra and the
exterior product rule%
\begin{equation}
\overset{p}{\rho}\mathbf{m}^{i}=(-1)^{p}\mathbf{m}^{i}\overset{p}{\rho};\text{
}\mathbf{m}^{i}\mathbf{m}^{j}=-\mathbf{m}^{j}\mathbf{m}^{i}.
\end{equation}
together with the condition of linear independence. \ For a given choice of
$\{\mathbf{m}^{i}\}$, a type $N=2$ generalized p-form, $\overset{p}{\mathbf{r}%
}$, can be written as%
\begin{equation}
\overset{p}{\mathbf{r}}=\overset{p}{\rho}+\overset{p+1}{\rho}_{i}%
\mathbf{m}^{i}+\overset{p+2}{\rho}\mathbf{m}^{1}\mathbf{m}^{2},
\end{equation}
where $\overset{p}{\rho},\overset{p+1}{\rho}_{i},\overset{p+1}{\rho}%
_{i},\overset{p+2}{\rho}$ are ordinary forms, respectively a $p-$form, two
($p+1)-$forms and a ($p+2)-$form. $\ $Hence, given a linearly independent pair
$\{\mathbf{m}^{i}\}$, $\overset{p}{\mathbf{r}}$ is determined by an ordered
quadruple of ordinary differential forms%
\begin{equation}
\overset{p}{\mathbf{r}}=(\overset{p}{\rho},\overset{p+1}{\rho}_{1}%
,\overset{p+1}{\rho}_{2},\overset{p+2}{\rho}).
\end{equation}

When it is assumed that the exterior derivative, $d,$ of generalized forms
satisfies the usual properties, in particular $d^{2}=0$, and that the exterior
derivative of any basis minus one form is a type $N=2$ generalized zero form,
that is
\[
d\mathbf{m}^{i}=\mu^{i}-\nu_{j}^{i}\mathbf{m}^{j}+\rho^{i}\mathbf{m}%
^{1}\mathbf{m}^{2}%
\]
where $\mu^{i}$, $\nu_{j}^{i}$ and $\rho^{i}$ are respectively zero- one- and
two-forms, it is a straightforward matter to show that the freedom in the
choice of basis minus one- forms,
\[
\mathbf{m}^{i}\mapsto(\Lambda^{-1})_{j}^{i}\mathbf{m}^{j}+\Upsilon
^{i}\mathbf{m}^{1}\mathbf{m}^{2},
\]
where the determinant of the matrix-valued function $\Lambda$ is non-zero and
$\Upsilon^{i}$ are one-forms, can be used to construct a basis of minus
one-forms satisfying
\begin{equation}
d\mathbf{m}^{i}=\epsilon^{i},
\end{equation}
where $\epsilon^{1}$ and $\epsilon^{2}$are constants, \cite{rob5}.

In this paper bases satisfying Eq.(4), with at least one of the constants
non-zero, will be used. \ \ It then follows that the exterior derivative of a
type $N=2$ generalized $p-$form $\overset{p}{\mathbf{r}}$ is the $(p+1)-$form%
\begin{align}
d\overset{p}{\mathbf{r}}  &  =d\overset{p}{\rho}+(-1)^{p+1}\rho_{i}%
\epsilon^{i}+(d\overset{p+1}{\rho}_{1}+(-1)^{p+1}\epsilon^{2}%
\overset{p+2}{\rho})\mathbf{m}^{1}+\\
&  +(d\overset{p+1}{\rho}_{2}+(-1)^{p}\epsilon^{1}\overset{p+2}{\rho
})\mathbf{m}^{2}+d\overset{p+2}{\rho}\mathbf{m}^{1}\mathbf{m}^{2}\nonumber
\end{align}
where $d$ is the ordinary exterior derivative when acting on ordinary forms.
\ The exterior derivative $d:$ $\Lambda_{(2)}^{p}(M)\rightarrow\Lambda
_{(2)}^{p+1}(M)$ is an anti-derivation of degree one,%
\begin{align}
d(\overset{p}{\mathbf{r}}\overset{q}{\mathbf{s}})  &
=(d\overset{p}{\mathbf{r}})\overset{q}{\mathbf{s}}+(-1)^{p}%
\overset{p}{\mathbf{r}}d\overset{q}{\mathbf{s}},\\
d^{2}  &  =0.\nonumber
\end{align}
and $(\Lambda_{(2)}^{\bullet}(M),d)$ is a differential graded algebra.

If $\varphi$ is a smooth map between manifolds $P$ and $M,$ $\varphi
:P\rightarrow M,$ then the induced map of type $N=2$ generalized forms,
$\varphi_{(2)}^{\ast}:\Lambda_{(2)}^{p}(M)\rightarrow\Lambda_{(2)}^{p}(P)$, is
the linear map defined by using the standard pull-back map, $\varphi^{\ast}$,
for ordinary forms%
\begin{equation}
\varphi_{(2)}^{\ast}(\overset{p}{\mathbf{r}})=\varphi^{\ast}(\overset{p}{\rho
})+\varphi^{\ast}(\overset{p+1}{\rho}_{i})\mathbf{m}^{i}+\varphi^{\ast
}(\overset{p+2}{\rho})\mathbf{m}^{1}\mathbf{m}^{2},
\end{equation}
and $\varphi_{(2)}^{\ast}(\overset{p}{\mathbf{r}}\overset{q}{\mathbf{s}%
})=\varphi_{(2)}^{\ast}(\overset{p}{\mathbf{r}})\varphi_{(2)}^{\ast
}(\overset{q}{\mathbf{s}})$. \ Hence $\varphi_{(2)}^{\ast}(\mathbf{m}%
^{i}\mathbf{)=m}^{i}$.

Integration is defined using polychains, \cite{rob6}. \ A $p-$polychain of
type $N=2$ in $M$, denoted $\mathbf{c}_{p}$ is an ordered quadruple of
ordinary (real, singular) chains in $M$%
\begin{equation}
\mathbf{c}_{p}=(c_{p},c_{p+1}^{1},c_{p+1}^{2},c_{p+2}),
\end{equation}
where $c_{p}$ is an ordinary $p-$chain, $c_{p+1}^{1}$ and $c_{p+1}^{2}$ are
ordinary $p+1-$chains and $c_{p+2}$ is and ordinary ordinary\ $p+2-$chain.
\ The ordinary chains boundaries are denoted by $\partial$ and the boundary of
the polychain $\mathbf{c}_{p}$ is the $(p-1)-$polychain $\partial
\mathbf{c}_{p}$ given by
\begin{equation}
(\partial c_{p},\partial c_{p+1}^{1}+(-1)^{p}\epsilon^{1}c_{p},\partial
c_{p+1}^{2}+(-1)^{p}\epsilon^{2}c_{p},\partial c_{p+2}+(-1)^{p}\epsilon
^{2}c_{p+1}^{1}+(-1)^{p-1}\epsilon^{1}c_{p+1}^{2}),
\end{equation}
and%
\begin{equation}
\partial^{2}\mathbf{c}_{p}=0.
\end{equation}

When $N=2$ the integral of a generalized form $\overset{p}{\mathbf{r}}$ over a
polychain $\mathbf{c}_{p}$ is%
\begin{equation}
\int_{\mathbf{c}_{p}}\overset{p}{\mathbf{r}}=\int_{c_{p}}(\overset{p}{\rho
}+\int_{c_{p+1}^{1}}\overset{p+1}{\rho}_{1}+\int_{c_{p+1}^{2}}%
\overset{p+1}{\rho}_{2}+\int_{c_{p+2}}\overset{p+2}{\rho}.
\end{equation}
and Stokes' theorem applies%
\begin{equation}
\int_{\mathbf{c}_{p}}d\overset{p-1}{\mathbf{r}}=\int_{\partial\mathbf{c}_{p}%
}\overset{p-1}{\mathbf{r}}.
\end{equation}

Under a change of basis minus one-forms $\mathbf{m}^{1}$ and $\mathbf{m}^{2}$%
\begin{equation}
\mathbf{m}^{i}\mapsto T_{j}^{i}\mathbf{m}^{j},
\end{equation}
where $\left(  T_{j}^{i}\right)  $ is a constant matrix and $T$= det$\left(
T_{j}^{i}\right)  $ is non-zero,%
\begin{equation}
\epsilon^{i}\mapsto T_{j}^{i}\epsilon^{j}.
\end{equation}
and the components of $\overset{p}{\mathbf{r}}$ transform as%
\begin{equation}
(\overset{p}{\rho},\overset{p+1}{\rho}_{1},\overset{p+1}{\rho}_{2}%
,\overset{p+2}{\rho})\mapsto(\overset{p}{\rho},(T^{-1})_{1}^{j}%
\overset{p+1}{\rho}_{j},(T^{-1})_{2}^{j}\overset{p+1}{\rho}_{j},(T^{-1}%
\overset{p+2}{\rho}).
\end{equation}
The form of the right hand side of Eq.(11) is then preserved if the components
of $\mathbf{c}_{p}$ transform as
\begin{equation}
(c_{p},c_{p+1}^{1},c_{p+1}^{2},c_{p+2})\mapsto(c_{p},(T)_{j}^{1}c_{p+1}%
^{j},(T)_{j}^{2}c_{p+1}^{j},Tc_{p+2}).
\end{equation}
In the following sections the usual definitions will be extended to admit
complex coefficients and the complex (and complex conjugate) combinations%
\begin{align}
\mathbf{m}  &  =\epsilon^{-1}(\mathbf{m}^{1}+i\mathbf{m}^{2}),\\
\overline{\mathbf{m}}  &  =\overline{\epsilon}^{-1}(\mathbf{m}^{1}%
-i\mathbf{m}^{2}),\nonumber\\
\epsilon &  =\epsilon^{1}+i\epsilon^{2},\nonumber
\end{align}
which satisfy%
\begin{equation}
d\mathbf{m=}d\overline{\mathbf{m}}=1,
\end{equation}
will be used.

Just as the algebra and differential calculus of ordinary differential forms
on $M$ can be expressed in terms of functions and vector fields on the reverse
parity tangent bundle, $\Pi TM$, of $M$, \cite{voronov}, \cite{witten3},
generalized forms can be represented in terms of functions and vector fields
on the Whitney sum of $\Pi TM$ and a trivial reverse parity line bundle over
$M$. \ For type $N=2$ forms the latter is a trivial bundle with fibre
$\mathbb{R}^{2}$ replaced by $\mathbb{R}^{0\mid2}$. \ Further details about
this and type $N$ generalized form-valued vector fields are in \cite{rob8}.

\section{Type N=2 generalized connections}

A generalized connection $\mathbf{A}$ with values in the Lie algebra,
$\mathfrak{g}$, of a matrix Lie group $G$ is defined in essentially the same
way as ordinary connections, as for example described in \cite{nak}, except
that ordinary forms are replaced by generalized forms. \ In this paper it will
be sufficient to use matrix representations of Lie groups and Lie algebras and
connections will be represented by matrix-valued generalized forms on $M$.
\ The primary focus will be on real generalized connection forms. $\mathbf{A}%
$, but generalization to complex generalized connection forms is trivial.

Under a gauge transformation by $g:M\rightarrow G$ a $\mathfrak{g}$-valued
generalized connection one-form transforms in the usual way%
\begin{equation}
\mathbf{A}\rightarrow(g^{-1})dg+(g^{-1})\mathbf{A}g
\end{equation}

The generalized curvature two-form $\mathbf{F}$ is the generalized two-form%
\begin{equation}
\mathbf{F=dA}+\mathbf{AA},
\end{equation}
Under the transformation in Eq.(19)%
\begin{equation}
\mathbf{F}\rightarrow(g^{-1})\mathbf{F}g.
\end{equation}
In terms of the complex basis introduced in Eq.(17) above a type $N=2$
connection one-form can be written as%
\begin{equation}
\mathbf{A}=\alpha-\phi\mathbf{m-}\overline{\phi}\overline{\mathbf{\mathbf{m}}%
}+i\chi\mathbf{m}\overline{\mathbf{\mathbf{m}}}\mathbf{,}%
\end{equation}
where for real $\mathbf{A}$, $\alpha$ and $\tau$ are, respectively, a real
ordinary $\mathfrak{g}$-valued one-form and three-form on $M$ and $\phi$ is an
ordinary complex $\mathfrak{g}$-valued two-form with complex conjugate
$\overline{\phi}$. \ The curvature two-form is
\begin{align}
\mathbf{F}  &  =\mathcal{F}-\phi\mathbf{-}\overline{\phi}\mathbf{-}%
(D\phi-i\chi)\mathbf{m-(}D\overline{\phi}+i\chi)\overline{\mathbf{m}}%
+(iD\chi+\phi\overline{\phi}-\overline{\phi}\phi)\mathbf{m}\overline
{\mathbf{m}}\mathbf{,}\\
\mathcal{F}  &  =d\alpha+\alpha\alpha,\nonumber\\
D\overset{p}{\rho}  &  =d\overset{p}{\rho}+\alpha\overset{p}{\rho}%
+(-1)^{p+1}\overset{p}{\rho}\alpha.\nonumber
\end{align}

The generalized curvature two-form is zero if and only if the generalized
connection can be written in the form%
\begin{equation}
\mathbf{A=}\alpha-\frac{1}{2}(\mathcal{F+}i\tau)\mathbf{m-}\frac{1}%
{2}(\mathcal{F-}i\tau)\overline{\mathbf{m}}+\frac{i}{2}D\tau\mathbf{m}%
\overline{\mathbf{m}},
\end{equation}

where, for real $\mathbf{A}$, $\tau$ is an arbitrary real $\mathfrak{g-}%
$valued two-form.

Henceforth in this paper it will be assumed, unless stated otherwise, that a
connection $\mathbf{A}$ has zero trace, $Tr\mathbf{A}=0$.

The generalized Chern-Pontrjagin class is determined by a generalized
four-form $\mathbf{CP}$%
\begin{equation}
\mathbf{CP}=\frac{1}{8\pi^{2}}Tr(\mathbf{FF}),
\end{equation}
\ which is equal to the exterior derivative of the generalized Chern-Simons
three-form $\mathbf{CS}$%
\begin{equation}
\mathbf{CS}=\frac{1}{8\pi^{2}}Tr(\mathbf{AF}-\frac{1}{3}\mathbf{AAA}).
\end{equation}
By Stokes' theorem, Eq.(12),\ for a polychain $\mathbf{c}_{4}$%
\begin{equation}
\int_{\mathbf{c}_{4}}\mathbf{CP}=\int_{\mathbf{\partial c}_{4}}\mathbf{CS}%
\end{equation}
and when $\mathbf{c}_{3}=\partial\mathbf{c}_{4}$ under the gauge
transformation given by Eq.(19)%
\[
\int_{\mathbf{c}_{3}}\mathbf{CS\rightarrow}\int_{\mathbf{c}_{3}}\mathbf{CS}.
\]

Using the generalized connection $\mathbf{A}$ given in Eq.(22)%
\[
\mathbf{CS}\mathbf{=}\Pi+\Delta\mathbf{m}+\overline{\Delta}\overline
{\mathbf{m}}+\Xi\mathbf{m}\overline{\mathbf{m}}%
\]
where%
\begin{align}
\Pi &  =\frac{1}{8\pi^{2}}Tr\{\alpha\mathcal{F}-\frac{1}{3}\alpha\alpha
\alpha-\alpha(\phi+\overline{\phi)}\},\\
\Delta &  =\frac{1}{8\pi^{2}}Tr\{\phi(\phi+\overline{\phi)})+d(\alpha
\phi)-2\phi\mathcal{F}+i\alpha\chi\},\nonumber\\
\overline{\Delta}  &  =\frac{1}{8\pi^{2}}Tr\{\overline{\phi}(\phi
+\overline{\phi)})+d(\alpha\overline{\phi})-2\overline{\phi}\mathcal{F}%
-i\alpha\chi\},\nonumber\\
\Xi &  =\frac{1}{8\pi^{2}}Tr\{\overline{\phi}D\phi-\phi D\overline{\phi
}+2i(\mathcal{F}-\phi-\overline{\phi})\chi\},\nonumber
\end{align}
and%
\begin{align}
\int_{\mathbf{\partial c}_{4}}\mathbf{CS}  &  \mathbf{=}\int_{\partial c_{4}%
}\Pi+\int_{c_{4}}(\Delta+\overline{\Delta})+\\
&  +(\epsilon\overline{\epsilon})^{-1}[\int_{\partial c_{5}^{1}}%
(\overline{\epsilon}\Delta+\epsilon\overline{\Delta})+i\int_{\partial
c_{5}^{2}}(\overline{\epsilon}\Delta-\epsilon\overline{\Delta})+2\int%
_{\partial c_{6}+\epsilon^{2}c_{5}^{1}-\epsilon^{1}c_{5}^{2}}\Xi].\nonumber
\end{align}
In the following section type $N=2$ Chern-Simons integrals for a boundary
polychain $\mathbf{c}_{3}=\partial\mathbf{c}_{4}$ will be used as action integrals.

\section{Connections, metrics and gravity}

In this section certain type $N=2$ generalized connections $\mathbf{A}$, on an
$n\geqq6$ dimensional manifold $M$ will be considered. \ The general formalism
follows that in \cite{rob9} where type $N=1$ forms were used. \ The
connections will be represented by a $5\times5$ matrix-valued generalized
one-forms%
\begin{equation}
\mathbf{A=}\left(
\begin{array}
[c]{cc}%
\mathbf{A}_{b}^{a} & -\sigma\mathbf{A}^{a}\\
\mathbf{A}_{b} & 0
\end{array}
\right)  .
\end{equation}
and take values in the Lie algebra, $\mathfrak{g}$, of $G$, where\ $G$ is
$SO(3+1,1)$ when $\sigma=1$; $SO(3,1+1)$ when $\sigma=-1$ and $ISO(3,1)$ when
$\sigma=0$. \ \ In the first two cases, which will be of primary interest
here, the metric is given by the $5\times5$ matrix%
\begin{equation}
\left(
\begin{array}
[c]{cc}%
\eta_{ab} & 0\\
0 & \sigma
\end{array}
\right)  ,
\end{equation}%
\[
\eta_{ab}=diag(-1,1,1,1)
\]
and $\mathbf{A}_{b}=\eta_{ba}\mathbf{A}^{a}$. \ Latin indices ranging and
summing from $1$ to $4$. \ The generalized curvature of $\mathbf{A}$ is given
by%
\begin{equation}
\mathbf{F}=d\mathbf{A}+\mathbf{AA}=\left(
\begin{array}
[c]{cc}%
\mathbf{F}_{b}^{a} & -\sigma\mathbf{F}^{a}\\
\mathbf{F}_{b} & 0
\end{array}
\right)  ,
\end{equation}
where%
\begin{align}
\mathbf{F}_{b}^{a}  &  =\mathbf{dA}_{b}^{a}+\mathbf{A}_{c}^{a}\mathbf{A}%
_{b}^{c}-\sigma\mathbf{A}^{a}\mathbf{A}_{b},\\
\mathbf{F}^{a}  &  =d\mathbf{A}^{a}+\mathbf{A}_{b}^{a}\mathbf{A}^{b},\text{
}\mathbf{F}_{b}=\eta_{bc}\mathbf{F}^{c}.\nonumber
\end{align}
Now let

\
\begin{align}
\mathbf{A}_{b}^{a}  &  =\omega_{b}^{a}-(^{-}\Omega_{b}^{a}-\frac{\sigma}%
{l^{2}}^{-}\Sigma_{b}^{a})\mathbf{m-(}^{+}\Omega_{b}^{a}\mathbf{-}\frac
{\sigma}{l^{2}}^{+}\Sigma_{b}^{a}\mathbf{)}\overline{\mathbf{\mathbf{m}}%
}\mathbf{,}\\
\mathbf{A}^{a}  &  =\frac{1}{l}\theta^{a}\mathbf{,}\nonumber
\end{align}
where $l$ is a non-zero constant, $\omega_{ab}=-\omega_{ba}$ are ordinary real
one-forms and $\theta^{a}$ are four real ordinary one-forms on $M$.
\ Furthermore if%
\begin{equation}
\Omega_{b}^{a}=d\omega_{b}^{a}+\omega_{c}^{a}\omega_{b}^{c}%
\end{equation}
denotes the curvature of $\omega_{b}^{a}$ (regarded as an ordinary connection)
then $^{\pm}\Omega_{ab}$ are respectively the $so(3,1)$ self-dual and
anti-self dual parts of $\Omega_{b}^{a}$. \ That is%
\begin{equation}
^{\pm}\Omega_{ab}=\frac{1}{2}(\Omega_{ab}\mp i^{\ast}\Omega_{ab}),
\end{equation}
where $^{\ast}\Omega_{ab}=\frac{1}{2}\varepsilon_{abcd}\Omega^{cd}$ and the
totally skew symmetric Levi-Civita symbol satisfies $\varepsilon_{1234}=1$.
\ Furthermore $\Sigma^{ab}=\theta^{a}\theta^{b}$ and $^{\pm}\Sigma^{ab}$
denote its $so(3,1)$ self and anti-self dual parts $^{\pm}\Sigma^{ab}=\frac
{1}{2}(\Sigma^{ab}\mp i^{\ast}\Sigma^{ab})$.

Then for this connection the curvature $\mathbf{F}$ in Eq.(32) is given by%
\begin{align}
\mathbf{F}_{b}^{a}  &  =\frac{\sigma}{l^{2}}(D^{-}\Sigma^{a}{}_{b}%
\mathbf{m}+D^{+}\Sigma^{a}{}_{b}\overline{\mathbf{m}})\mathbf{,}\\
\mathbf{F}^{a}  &  =\frac{1}{l}[D\theta^{a}+(^{-}\Omega_{b}^{a}-\frac{\sigma
}{l^{2}}^{-}\Sigma_{b}^{a})\theta^{b}\mathbf{m+\mathbf{(^{+}}}\Omega_{b}%
^{a}\mathbf{\mathbf{-}}\frac{\sigma}{l^{2}}^{+}\Sigma_{b}^{a}%
\mathbf{\mathbf{)}}\theta^{b}\mathbf{\overline{\mathbf{\mathbf{m}}}%
]},\nonumber
\end{align}
Here the covariant exterior derivative with respect to $\omega_{b}^{a}$ is
denoted $D$ so that%
\begin{align}
D\theta^{a}  &  =d\theta^{a}+\omega_{b}^{a}\theta^{b},\\
D\text{ }^{\pm}\Sigma_{b}^{a}  &  =d^{\pm}\Sigma_{b}^{a}+\omega_{c}^{a}\text{
}^{\pm}\Sigma_{b}^{c}-\text{ }^{\pm}\Sigma_{c}^{a}\omega_{b}^{c}\nonumber\\
&  =d^{\pm}\Sigma_{b}^{a}+^{\pm}\omega_{c}^{a}\text{ }^{\pm}\Sigma_{b}%
^{c}-^{\pm}\Sigma_{c}^{a}\text{ }^{\pm}\omega_{b}^{c},\nonumber
\end{align}
where $^{\pm}\omega_{b}^{a}=\frac{1}{2}(\omega^{ab}\mp i^{\ast}\omega^{ab})$
respectively denote the $so(3,1)$ self-dual and anti-self dual parts of
$\omega_{b}^{a}.$

The generalized connection $\mathbf{A}$ is flat if and only if $\mathbf{F}=0$
and then%
\begin{align}
D\theta^{a}  &  =0,\\
(^{-}\Omega_{b}^{a}-\frac{\sigma}{l^{2}}\text{ }^{-}\Sigma_{b}^{a})\theta^{b}
&  =0,\nonumber\\
(^{+}\Omega_{b}^{a}-\frac{\sigma}{l^{2}}\text{ }^{+}\Sigma_{b}^{a})\theta^{b}
&  =0.\nonumber
\end{align}

Suppose now that when the $4$ ordinary one-forms $\{\theta^{a}\}$ are pulled
back to a four dimensional sub-manifold $S\subseteq M$ they are linearly
independent and so form an orthonormal basis for a Lorentzian metric
$ds^{2}=\eta_{ab}\theta^{a}\otimes\theta^{b}$ there. \ Then when Eq.(39)\ is
satisfied it follows that this metric satisfies Einstein's vacuum field
equations on $S$ with cosmological constant $\Lambda=\frac{3\sigma}{l^{2}}$.

In passing it is interesting to note the sub-case of the complex connection
$_{asd}\mathbf{A}_{\text{ }}$given by Eq.(30) with $\sigma=0$ and
\begin{equation}
_{asd}\mathbf{A}_{b}^{a}=\text{ }^{-}\omega_{b}^{a}-\text{ }^{-}\Omega_{b}%
^{a}\mathbf{m};\text{ }_{asd}\mathbf{A}_{a}=\theta_{a};\text{ }^{-}\Omega
_{b}^{a}=d\text{ }^{-}\omega_{b}^{a}+\text{ }^{-}\omega_{c}^{a}\text{ }%
^{-}\omega_{b}^{c}%
\end{equation}
with curvature $_{asd}\mathbf{F}$ given by%
\begin{align}
_{asd}\mathbf{F}_{b}^{a}  &  =0,\\
_{asd}\mathbf{F}_{a}  &  =d\theta_{a}+\text{ }^{-}\omega_{ab}\theta^{b}+\text{
}^{-}\Omega_{ab}\theta^{b}\mathbf{m,}\nonumber
\end{align}

If this generalized connection is flat the (complex) metric, $ds^{2}=\eta
_{ab}\theta^{a}\otimes\theta^{b}$, determined analogously to the real
four-metric above, is half flat with anti-self dual curvature $^{-}\Omega
_{b}^{a}$.

For the connection and curvature given by Eqs.(30), (32) and (33) the
generalized Chern-Pontrjagin and Chern-Simons forms are%
\begin{align}
\mathbf{CP}  &  \mathbf{=}\frac{1}{8\pi^{2}}[\mathbf{F}_{b}^{a}\mathbf{F}%
_{a}^{b}-2\sigma\mathbf{F}^{a}\mathbf{F}_{a}]\\
\mathbf{CS}  &  \mathbf{=}\frac{1}{8\pi^{2}}[\mathbf{A}_{b}^{a}\mathbf{F}%
_{a}^{b}-\frac{1}{3}\mathbf{A}_{b}^{a}\mathbf{A}_{c}^{b}\mathbf{A}_{a}%
^{c}-2\sigma\mathbf{A}^{a}\mathbf{F}_{a}+\sigma\mathbf{A}_{b}^{a}%
\mathbf{A}^{b}\mathbf{A}_{a}].\nonumber
\end{align}

Computation, using Eqs.(34) and (37), gives%
\begin{align}
\mathbf{CS}  &  \mathbf{=}\frac{1}{8\pi^{2}}\{-\frac{1}{3}\omega_{b}^{a}%
\omega_{c}^{b}\omega_{a}^{c}+\frac{\sigma}{l^{2}}\omega_{b}^{a}\theta
^{b}\theta_{a}-\frac{2\sigma}{l^{2}}\theta_{a}D\theta^{a}+\\
&  +[\frac{1}{3}d(\text{ }^{-}\omega_{.b}^{a}\text{ }^{-}\omega_{c}^{b}\text{
}^{-}\omega_{a}^{c}-3\frac{\sigma}{l^{2}}\text{ }^{-}\omega_{b}^{a}\text{
}^{-}\Sigma_{.a}^{b})-\frac{2\sigma}{l^{2}}\text{ }^{-}\Omega_{ab}\Sigma
^{ab}+\frac{i\sigma^{2}}{4l^{4}}\varepsilon_{abcd}\theta^{a}\theta^{b}%
\theta^{c}\theta^{d}]\mathbf{m+}\nonumber\\
&  +[\frac{1}{3}d(\text{ }^{+}\omega_{b}^{a}\text{ }^{+}\omega_{c}^{b}\text{
}^{+}\omega_{a}^{c}-3\frac{\sigma}{l^{2}}\text{ }^{+}\omega_{b}^{a}\text{
}^{+}\Sigma_{.a}^{b})-\frac{2\sigma}{l^{2}}\text{ }^{+}\Omega_{ab}\Sigma
^{ab}+\frac{i\sigma^{2}}{4l^{4}}\varepsilon_{abcd}\theta^{a}\theta^{b}%
\theta^{c}\theta^{d}]\overline{\mathbf{m}}\mathbf{\}.}\nonumber
\end{align}
Using Eq.(17) and integrating over a boundary polychain $\mathbf{c}%
_{3}=\partial\mathbf{c}_{4},$ as in Eq.(9) with $p=4,$ gives%
\begin{align}
\int_{\mathbf{c}_{3}}\mathbf{CS}  &  \mathbf{=}\frac{\sigma}{4\pi^{2}l^{2}%
}\{-\int_{\partial c_{4}}\theta^{a}D\theta_{a}-\int_{c_{4}}\Omega_{ab}%
\Sigma^{a}{}^{b}+\\
&  -\int_{\partial c_{5}^{1}}[i\kappa^{2}(\text{ }^{+}\Omega_{ab}\Sigma
^{ab}-\text{ }^{-}\Omega_{ab}\Sigma^{ab}-\frac{\sigma}{4l^{2}}\varepsilon
_{abcd}\theta^{a}\theta^{b}\theta^{c}\theta^{d})+\kappa^{1}\Omega_{ab}%
\Sigma^{ab}]\nonumber\\
&  +\int_{\partial c_{5}^{2}}[i\kappa^{1}(\text{ }^{+}\Omega_{ab}\Sigma
^{ab}-\text{ }^{-}\Omega_{ab}\Sigma^{ab}-\frac{\sigma}{4l^{2}}\varepsilon
_{abcd}\theta^{a}\theta^{b}\theta^{c}\theta^{d})-\kappa^{2}\Omega_{ab}%
\Sigma^{ab}]\}.\nonumber
\end{align}
where%
\[
\kappa^{1}=\frac{\epsilon^{1}}{\epsilon\overline{\epsilon}};\text{ }\kappa
^{2}=\frac{\epsilon^{2}}{\epsilon\overline{\epsilon}}.
\]

Now consider this expression as a generalized Chern-Simons action integral and
the case where the four ordinary one-forms $\{\theta^{a}\}$ are linearly
independent on the four dimensional sub-manifolds (chains), $c_{4},\partial
c_{5}^{1}$and $\partial c_{5}^{2}$. \ Since the one forms constitute an
orthonormal basis there for a Lorentzian metric $ds^{2}=\eta_{ab}\theta
^{a}\otimes\theta^{b}$ this action can now be rewritten as%
\begin{align}
\int_{\mathbf{c}_{3}}\mathbf{CS}  &  \mathbf{=}\frac{\sigma}{4\pi^{2}l^{2}%
}\{-\int_{\partial c_{4}}\theta^{a}D\theta_{a}-\int_{c_{4}}\Omega_{ab}%
\theta^{a}\theta^{b}+\\
&  -\int_{\partial c_{5}^{1}}[\kappa^{2}(R-2\Lambda)V+\kappa^{1}\Omega
_{ab}\Sigma^{ab}]+\nonumber\\
&  +\int_{\partial c_{5}^{2}}[\kappa^{1}(R-2\Lambda)V-\kappa^{2}\Omega
_{ab}\Sigma^{ab}]\},\nonumber
\end{align}
where on $\partial c_{5}^{1}$and $\partial c_{5}^{2}$%
\begin{align}
\Omega_{b}^{a}  &  =\frac{1}{2}R_{bcd}^{a}\theta^{c}\theta^{d},R=\eta
^{bd}R_{bad}^{a},\\
V  &  =\theta^{1}\theta^{2}\theta^{3}\theta^{4},\Lambda=\frac{3\sigma}{l^{2}%
}\nonumber
\end{align}
and the action terms there correspond to the usual first order (Palatini)
Einstein-Hilbert action terms augmented by the term proportional to
$\Omega_{ab}\Sigma^{ab}$, \cite{holst} \ 

On the boundary manifolds $\partial c_{5}^{1}$and $\partial c_{5}^{2}$ the
Euler-Lagrange equations are Einstein's \ vacuum field equations,
($G_{ab}+\Lambda\eta_{ab})\theta^{a}\otimes\theta^{b}=0$, with cosmological
constant $\Lambda=\frac{3\sigma}{l^{2}}$. \ In addition the variation of the
first two terms gives%
\begin{equation}
\delta(-\int_{\partial c_{4}}\theta^{a}D\theta_{a}-\int_{c_{4}}\Omega
_{ab}\theta^{a}\theta^{b})=-2\int_{\partial c_{4}}\delta\theta^{a}D\theta
_{a}+\int_{c_{4}}2\delta\theta^{b}\Omega_{ab}\theta^{a}-\delta\omega_{b}%
^{a}D(\theta^{a}\theta_{b}).
\end{equation}
When the geometry of\ the submanifolds is specified in more detail these
results can be interpreted more completely.

In conclusion it should be noted that the use of an anti-deSitter/deSitter
connection, invariant only under the Lorentz group, in an action principle
with the Einstein field equations as Euler Lagrange equations dates from the
late 1970's, \cite{macdowell}, \cite{stelle1} and\cite{stelle2}. \ However the
approach, initiated in these papers, which has recently been interpreted in
terms of Cartan geometries, \cite{wise}, is different from the one taken
here.

\end{document}